\documentstyle{l-aa}
\font\sixrm=cmr6
\newcommand{\magpt}[2]{\mbox{$\rm #1\hspace{-0.25em}\stackrel{m}{.}
      \hspace{-1.0mm}#2$}}                             
\newcommand{\dg}{$^{\circ}$}                  
\newcommand\ion[2]{\hbox{#1\,{\sixrm #2}}}
\newcommand\teff{$ {\rm T_{eff}}$}
\newcommand\logt{$\log {\rm T_{eff}}$}
\newcommand\logg{$\log {\rm g}$}
\newcommand{\Msolar}{\mbox{\,$\rm M_{\odot}$}}        
\newcommand{\Zsolar}{\mbox{\,$\rm Z_{\odot}$}}        
\begin{document}
\thesaurus{5(08.05.1; 08.08.2; 08.19.2; 08.16.3; 10.07.3 NGC~6752)}
\title{Hot HB stars in globular clusters - physical parameters and consequences
for theory}
\subtitle{III. NGC 6752 and its long blue vertical branch
\thanks{Based on observations collected at the European Southern Observatory}}

\author{S. Moehler
\inst{1}$^,$\inst{2}
 \and U. Heber \inst{3} \and G. Rupprecht \inst{4}}
\offprints{U. Heber}
\institute {Landessternwarte, K\"onigstuhl, 69117 Heidelberg,
Germany
\and Space Telescope Science Institute, 3700 San Martin Drive, Baltimore,
MD 21218, USA (e-mail: smoehler@stsci.edu)
\and Dr. Remeis-Sternwarte, Sternwartstr. 7, 96049 Bamberg, Germany
\and European Southern Observatory, Karl-Schwarzschild-Str. 2, 85748 Garching
bei M\"unchen, Germany}
\date{Received March 5, 1996; revised June 11,1996; accepted }
\maketitle
\begin{abstract}

We present spectroscopic analyses of 17 faint blue stars in the globular
cluster
NGC~6752, using optical and UV spectrophotometric data and intermediate
resolution optical spectra. Effective temperatures, surface gravities, and
helium abundances
of the stars are determined and compared to theoretical predictions. All
stars are helium deficient by factors ranging from 3 to more than 100,
indicative of gravitational settling of helium.
Stars with effective temperatures above about 20000~K (sdBs)
fit well
to the evolutionary tracks, whereas the cooler stars show lower surface
gravities than theoretically expected. This agrees with earlier findings
by Moehler et al. (\cite{mohe95}) for the faint blue stars in M~15.
Deriving masses from the atmospheric parameters and the cluster distance
leads to a mean mass of the
sdB stars of 0.50~\Msolar\ and a standard deviation of about
0.043 dex, which is below the value derived of the observational
errors and therefore consistent with a very narrow mass distribution.
The mean mass is in good agreement with
the (0.49 $\pm$ 0.02)~\Msolar\  expected if sdB stars are extreme horizontal
branch stars with a helium core of 0.48~\Msolar\ and an extremely thin
hydrogen layer. The cooler stars however show a significantly lower mean
{ mass of 0.30~\Msolar\ with a standard deviation of the mean of 0.059 dex}
in agreement with the findings of
Moehler et al. (\cite{mohe95}) and de Boer et al. (\cite{dbsc95}) for
the BHB stars in M~15 and NGC~6397, respectively. The problems presented by
the different mass distributions of BHB and sdB stars within one cluster
are discussed.

\end{abstract}
\section{Introduction}
The horizontal branches (HBs) in colour-magnitude diagrams (CMDs) of
galactic globular clusters show a large variety in morphology:
On the blue side of the RR~Lyrae gap one may find HBs extending down to very
faint magnitudes at the same blue colour (e.g. M~5,
Buonanno et al., \cite{buco81})
while others are short and limited to the region near M$_V$ = \magpt{0}{0}
(e.g. NGC 6397, Alcaino \& Liller, \cite{alli80}).
Also, some of these HBs appear to have gaps
(e.g. NGC 288, Buonanno et al., \cite{buco84}; M~15, Buoanno et al. ,
\cite{buco85})
or, at least, are not populated in a smooth manner (e.g. NGC~6752, Buonanno
et al., \cite{buca86}).

To distinguish the long and almost vertical extensions from the main body
of the horizontal branch, the term ``extended horizontal branch'' (EHB)
in contrast to ``blue horizontal branch'' (BHB) is used.
In the field of the
 Milky Way the EHB stars are identified with the subdwarf B (sdB)
 stars, which have \teff\ $>$~20000~K and \logg\ $>$~5 (Greenstein \&
Sargent, \cite{grsa74}; Heber et al., \cite{hebe84};
Moehler et al. \cite{mohe90}; Saffer et al., \cite{saff94}).

Two basic scenarios for the origin of the sdB stars have been put forward
in the literature: sdBs are assumed to have either originated as single
stars, which for some reason have lost more of their hydrogen envelope by
the time they arrive on the HB than most stars do, or they are the product
of some kind of binary interaction (see Bailyn et al. \cite{baea92}, for a
detailed discussion). In the single star scenario, the sdB mass is close to
0.5 \Msolar\ and determined by the canonical helium core mass at the core
helium flash (see Sweigart, \cite{swei94}).
The importance of binary evolution is supported by the high binary frequency
observed amongst the field sdB stars (Allard et al., \cite{alla94};
Theissen et al., \cite{thmo95}).
There are three variants of the binary scenario:

\begin{itemize}
\item [(I)] Common envelope evolution can lead to the formation of subluminous
O and B stars if Roche lobe overflow occurs during the ascend of the first
(case B) or second (case C) giant branch (Iben \& Tutukov \cite{ibtu85},
\cite{ibtu93}; Iben \cite{iben86}; Iben \& Livio \cite{ibli93}).
{ Well studied examples are HD\,128220 (sdO+ FIV) and AA\,Dor.
For case B evolution the sdB masses are
somewhat smaller than the canonical 0.5\,\Msolar\ (e.g. AA\,Dor,
M~=~0.3\,\Msolar,
Kudritzki et al., \cite{kudr82}), while for case C they are slightly larger
than the canonical value (e.g. HD\,128220, M~$>$~0.55\Msolar, Howarth \& Heber,
\cite{hohe90}).}

\item [(II)] Mengel et al. (\cite{meng76}) explored the possibility that
sdB stars could evolve from binaries in which mass transfer is initiated at
the tip of the giant branch, i.e. shortly before the ignition of the core
helium flash, resulting in sdB masses very close to the canonical 0.5\Msolar.

\item [(III)] sdB stars can also result from the merger of a double helium
white dwarf system  (Iben \& Tutukov, \cite{ibtu84}; Iben \cite{iben90})
with masses ranging from 0.3\,\Msolar\ to 0.9\,\Msolar\ and low mass stars
formed preferably. Bailyn \& Iben (\cite{baib89}) have explored this
possibility for globular cluster sdBs and argue that a few tens of sdBs can
be formed in a typical globular cluster, because binary-single star
interactions increase the merger rate considerably, especially for the
central regions of the cluster.

\end{itemize}
Therefore at least four scenarios for the origin of sdB stars are at hand.
They can be distinguished observationally if the mass distribution of sdB
stars can be determined, since the four scenarios predict quite different
mass distributions: the single star and Mengel's binary scenario (II)
predict a sharp mass distribution peaked at the canonical mass of
0.5\,\Msolar, while binary scenarios (I) and (III) predict broader
distributions peaked at masses below 0.5\,\Msolar. In order to find out
which is the dominant sdB production channel in globular clusters, we have
started to determine spectroscopic masses of blue horizontal branch stars
in the globular clusters M~15 (Moehler et al., \cite{mohe95}, paper~I),
NGC~6397 (de Boer et al.,\cite{dbsc95}, paper~II), and NGC~6752 (Heber et
al., \cite{heku86} and this paper)
from atmospheric parameters and
the known cluster distances.
The previous analyses of the BHB and EHB stars in NGC\,6752, M~15 and NGC~6397
showed
that
 \begin{enumerate}
\item Three stars in NGC\,6752 are bona fide EHB stars and cannot be
distinguished from field sdB stars. One very blue star is a sdOB star which
already evolved from the EHB.
\item The stars below the gap in M~15 are still BHB stars according to their
atmospheric  parameters \teff\ and \logg .
\item The BHB stars in M~15 and NGC~6397 show masses that are significantly
lower than predicted by canonical theory.
\end{enumerate}
NGC 6752 is up to now the only
cluster where the gap seen in the CMD really separates sdB stars from BHB
stars.
For a long time it has also been the only cluster where sdBs have been securely
 identified at all.
Additional sdB/sdOB candidates have been verified recently in M~15
(Durrell \& Harris, \cite{duha93}; Moehler \cite{moeh95})
and M~22 (Moehler et al., \cite{aas96}).

NGC 6752 has a large population of EHB stars and is therefore very well
suited for our aims. Early CMDs (e.g. Cannon,
\cite{cann81}) display a pronounced gap (between V~$\approx$~\magpt{16}{0}
and V~$\approx$~\magpt{17}{0}) separating the EHB (V~$>$~\magpt{17}{0}) from
the
BHB (V~$<$~\magpt{16}{0}). More recent CMDs (Buonanno et al. \cite{buca86}),
however, suggested that the gap is a region of the HB which is sparsely
populated rather than devoid of stars. Therefore, we aim to study more bona
fide EHB stars (below the gap, V~$<$~\magpt{17}{0}) to determine their masses
spectroscopically using the same methods as in Heber et al. (\cite{heku86})
and paper I. The second aim is to clarify the nature of the stars inside
the HB gap region (i.e. \magpt{16}{0}~$<$~V~$<$~\magpt{17}{0}), discovered by
Buonanno et al. (\cite{buca86}). We also include two stars brighter than
the HB gap because of their unusually blue colour.

\section{Observations}

\begin{figure}
\includegraphics{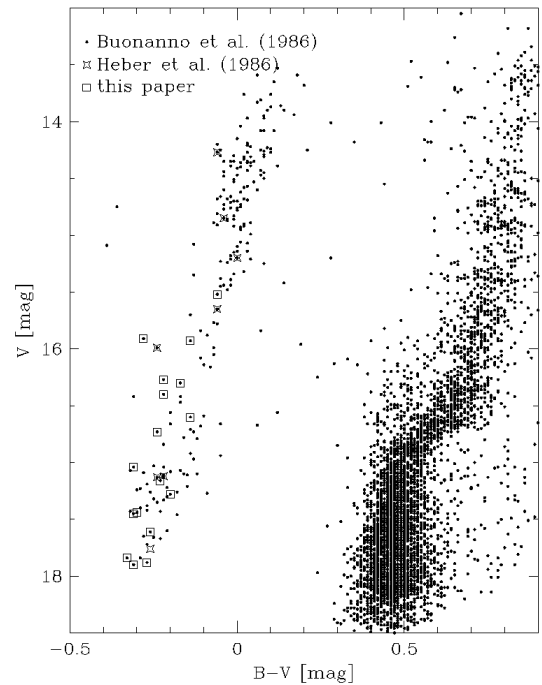}
\vspace{10.0cm}
\caption[]{The colour-magnitude diagram of NGC 6752 as published by
Buonanno et al. (1986). Stars analysed in this paper are marked by open
squares, open stars mark the objects discussed by Heber et al. (1986).
The star 3-188 of Heber et al. is not in the Buonanno data.
}
\end{figure}

We selected 17 blue stars from Buonanno et al. (\cite{buca86}) for
spectroscopic follow-up spectroscopy. { These stars are marked in Fig.~1
with squares. Open stars in this figure mark the stars discussed by
Heber et al. (\cite{heku86})}. Nine stars lie below the gap in the
CMD (V~$>$~\magpt{17}{0}), five stars are in the gap region
(\magpt{16}{0}~$<$~V~$<$~\magpt{17}{0}) and three stars are above the gap
(V~$<$~\magpt{16}{0}), two of which have very blue colours.
All optical spectra were obtained at telescopes of the ESO La Silla observatory
using focal reducers.
For calibration purposes we always observed 10 bias frames each night
and 5-10 flat-fields with a mean exposure level of about 10000 ADU each.
We obtained low resolution spectrophotometric data with large slit widths
to analyse the
flux distribution and medium resolution spectra to measure Balmer line
profiles and helium line equivalent widths.

The atmospheric extinction was corrected using the data of T\"ug (1977)
and the data for the flux standard stars were taken from Hamuy et al.
(\cite{hamu92}).

\begin{table*}
\begin{tabular}{|l|ll|r|rr|rr|}
\hline
Instrument/& Date & Telescope & CCD & pixel & no. of & gain & read-out \\
Mode & & & & size & pixels &  & noise \\
     & & & & [$\mu$m] & & [e$^-$/ADU] & [e$^-$]\\
\hline
EFOSC1 & 1992/07/04-07 & 3.6m & RCA \# 8 & 15 & 656 $\times$ 1024 & 2.1 & 33 \\
\hline
EMMI/B & 1993/07/23-26 & NTT & Tek \# 31 & 24 & 1024 $\times$ 1024 & 3.4 &
5.7\\
EMMI/R & &  & FA \# 34 & 15 & 2048 $\times$ 2048 & 1.5 & 6.6  \\
\hline
EFOSC2 & 1995/06/26-28 & 2.2m MPI/ESO & Th \# 19 & 19 & 1024 $\times$ 1024 &
2.1 & 4.3 \\
\hline
\end{tabular}
\vspace{3mm}
\begin{tabular}{|l|l|rrl|rr|}
\hline
Instrument/& Slit & Grism/ & Dispersion & Wavelength &
\multicolumn{2}{c|}{wavelength calibration}\\
Mode & width & Grating & & range & number of & r.m.s.\\
 & & & &  & lines used & error \\
 & [\arcsec] & &  [\AA/mm] & [\AA ] & & [\AA ] \\
\hline
EFOSC1/low & 5.0 & UV300 & 210 & 3200 -- 6400 & 11 & 0.5 \\
EFOSC1/med & 0.75 & B150 & 120 & 3800 -- 5600 & 12 & 0.3 \\
\hline
EMMI/DIMD B & 5.0 & \# 4 & 72 & 3400 -- 5300  & 7 & 0.8 \\
EMMI/BLMD B & 1.0 & \# 4 & 72 & 3200 -- 5100 & 15 & 0.3 \\
EMMI/DIMD R & 5.0 & \# 13 & 224 & 4000 -- 8500 & 14 & 0.5 \\
\hline
EFOSC2 & 5.0 & \# 1 & 442 & 1000 -- 9400 (nominal) & 16 & 1.7 \\
 & & & & 3400 -- 9400 (used) & & \\
\hline
\end{tabular}
\caption{Observation and reduction parameters}
\end{table*}

\subsection{Observations in 1992}

In 1992 we used EFOSC1 at the 3.6m telescope with a seeing between 1.5\arcsec\
and 2\arcsec.
Details of the setup are given in Table~1.
Since the RCA chip showed some column structure we took flat-fields using U and
B filters to achieve mean count levels between 5 and 10000 counts to allow
a correction of the column structure.
As flux standard star we observed Feige 110.
{ During these observations the slit angle was kept at its default
east-west orientation.}

\subsection{Observations in 1993}

In 1993 we observed with EMMI at the NTT, using the two-channel mode for
the low resolution spectrophotometric observations to get a longer wavelength
base.
Seeing values for these observations varied between 0.8\arcsec and 1.5\arcsec .
Since no EMMI grism allows observations below 3600~\AA\ (necessary to measure
the Balmer jump) we used a grating in the blue channel
and reduced the dispersion by binning  along the dispersion
axis by a factor of 3.
 For the intermediate resolution spectra we used only the blue channel
and did not bin.
{ the slit was kept at ist default east-west orientation (= 0\dg) except
for the spectrophotometric observations of
B~2395 (75\dg ), B~4009 (75\dg ), and B~4548 (85\dg ).
Positive angles mean anti-clockwise rotation.
Since we used a narrow slit for the medium resolution observations
and did not try to correct for atmospheric dispersion
 there are some slit losses at the blue end
of the medium resolution spectra.}
As flux standard stars we used LTT~6248, LTT~9491, and Feige~110.

\subsection {Observations in 1995}

In 1995 we used EFOSC2 at the 2.2m MPI/ESO telescope to re-observe the
low-resolution spectrophotometric data obtained with EMMI (see Section 3.2).
Hence we only used the low resolution setup.
Here we had seeing values between 1\arcsec\ and 1.5\arcsec .
Unfortunately the two nights we had were not really photometric.
As we had observed the standard stars quite frequently we could however
check the photometric conditions from the response curves.
As flux standard stars we used EG~274 and Feige~110.
{ During these observations we did not rotate the slit, but kept it at its
default east-west orientation.}

\subsection{IUE data}

In addition to the observations described above low resolution IUE SWP
spectra are available from the archive for the stars B~491, B~916, B~4009,
and B~4548. In addition to the standard reduction and calibration the
correction for the new white dwarf scale as described by Bohlin et al.
(\cite{bohl90}) and Bohlin (\cite{bohl96}) was performed.
{ we also recalibrated the old IUE data used in Heber et al.
(\cite{heku86}).
}

\begin{table}
\begin{tabular}{|l|rr|cc|cc|c|c|}
 \hline
Star & X & Y & \multicolumn{2}{|c|}{1992} &
\multicolumn{2}{|c|}{ 1993} & 1995
& v$_{\rm hel.}$\\
 & [\arcsec] & [\arcsec] & l & m & l & m & l & [km/s]\\
 \hline
B~210 & +432 & $-$389 & $\times$ & $\times$ & &  & &$-$13\\
B~491$^*$ & +344 & $-$256 & & & $\times$ & $\times$ & $\times$ & +7 \\
B~617 & +313 & $-$151 & $\times$ & $\times$ & $\times$ & & $\otimes$ & $-$28\\
B~852 & +275 & $-$98 & $\times$ & & & $\times$ & & 0\\
B~916$^*$ & +266 & $-$296 & & & $\times$ & $\times$ & $\otimes$ & +12\\
B~1288 & +212 & $-$290 & $\times$ & $\times$ & $\times$ & & $\times$ & $-$27\\
B~1509 & +174 & +264 & $\times$ & $\times$ & & $\times$ & & $-$33\\
B~1628 & +149 &  $-$41 & $\times$ & & & $\times$ & & $-$21\\
B~2126 &  +84 & +383 & $\times$ & $\times$ & $\times$ & & & $-$25\\
B~2162 &  +76 & $-$247 & & & $\times$ & $\times$ & $\otimes$ &$-$36 \\
B~2395 &  +41 & $-$277 & & & $\times$ & $\times$ & $\times$ & $-$52\\
B~3655 & $-$162 &  +88 & $\times$ & $\times$ & $\times$ & & & $-$19\\
B~3915 & $-$207 & +273 & & & & $\times$ & & $-$9\\
B~3975 & $-$218 & $-$246 & & & $\times$ & $\times$ & &+18\\
B~4009$^*$ & $-$222 & $-$532 & & & $\times$ & $\times$ & & $-$56\\
B~4380 & $-$276 & $-$135 & $\times$ & $\times$ & & & &$-$30\\
B~4548$^*$ & $-$307 & +444 & & & $\times$ & $\times$ & $\otimes$ & $-$65\\
\hline
\end{tabular}
\caption{List of observed objects. The star numbers refer to Buonanno et al.
(1986). Comparing the relative coordinates given there to other
photometry of NGC~6752,
we found that X$_B$ = -Y and vice versa.
For stars marked with $^*$ IUE SWP are available.
l (m) stands for low (medium) resolution spectra.
$\times$ marks spectra observed with the respective setup,
$\otimes$ marks low-resolution spectra obtained under
non-photometric conditions.
v$_{\rm hel.}$ is the heliocentric velocity.
}
\end{table}

\section{Data Reduction}

We describe the reduction of our data sets in some detail here since we
encountered several problems, esp. with the spectrophotometric data,
during the reduction and calibration of our observations. Since these problems
on one hand affect the quality of our data (and in consequence the
reliability of our
analysis) and were on the other hand produced by instruments much used by
observers at ESO we think that a detailed report is adequate.

\subsection{EFOSC1 and EMMI data}

We always averaged bias and flat-field frames for each night separately,
since there were small variations in the bias frames (up to a few counts/pixel)
and also a slight variation in the fringe patterns of the flat fields from
one night to the next (below 5\%).
The dark currents were determined from several long dark frames and turned
out to be negligible in all cases.
To correct the electronic offset we scaled the bias frames by the mean overscan
value of the science frame.
To take the spectral signature out of the averaged flat field frames
we fitted a 5$^{\rm th}$ (7$^{\rm th}$) order polynomial to the blue
(red) channel EMMI data.
The EFOSC1 flat fields could not be fitted well enough by polynomials. We
therefore averaged the mean flat-field of each night along the spatial
axis, smoothed it heavily to erase all small scale structure and divided
the original mean flat by the smoothed average.
For the EFOSC1 data we also had to correct column structure. This correction
was
performed as described in paper~I.

\begin{figure*}
\includegraphics{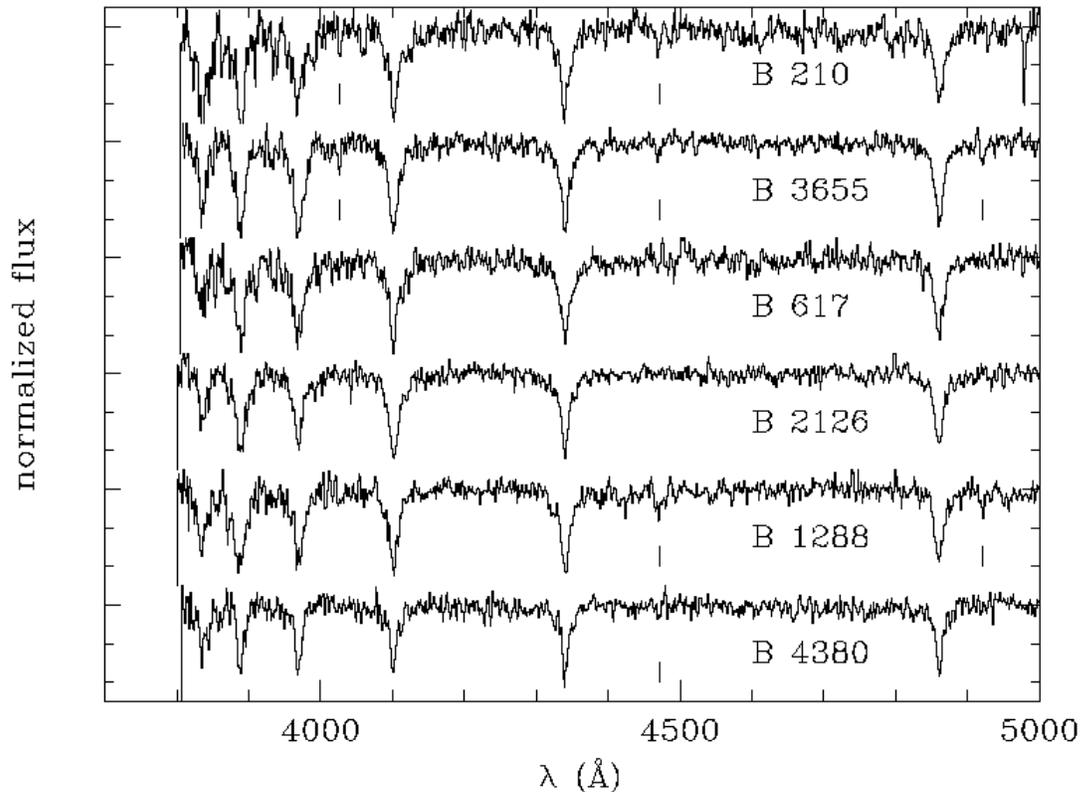}
\vspace{12.7cm}
\caption[]{Normalized medium resolution EFOSC1 spectra. The part shortward of
3900~\AA\ was normalized by taking the highest flux point as continuum
value. The spectra are sorted along decreasing Balmer line depths.
Small tickmarks mark steps of 20\%.}
\end{figure*}

\begin{figure*}
\includegraphics{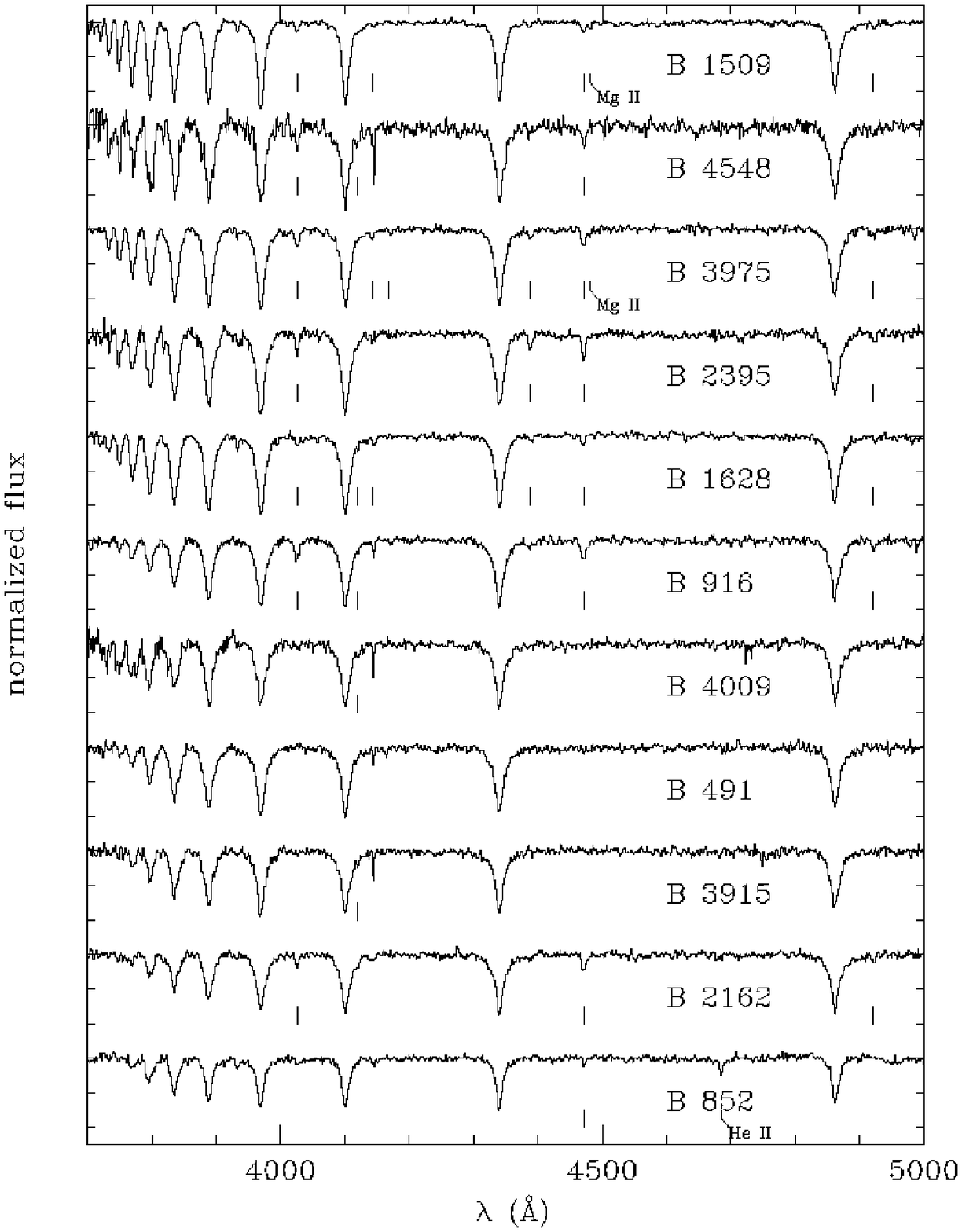}
\vspace{21cm}
\caption[]{Normalized medium resolution EMMI spectra. The part shortward of
3900~\AA\ was normalized by taking the highest flux point as continuum
value. The spectra are sorted along decreasing Balmer line depths.
Small tickmarks mark steps of 20\% .}
\end{figure*}

For the wavelength calibration we fitted
3$^{\rm rd}$ order polynomials to the dispersion relations.
We rebinned the frames two-dimensionally to constant wavelength steps.
Before the sky fit the frames were smoothed along the spatial axis to erase
cosmics in the background. To determine the sky background we had to find
regions without any stellar spectra, which were sometimes not close to the
place of the object's spectrum. Nevertheless the flat field correction
and wavelength calibration turned
out to be good enough that a linear (EFOSC1) or constant (EMMI)
fit to the spatial distribution of the sky light allowed to subtract the
sky background
at the object's position with sufficient accuracy. This means in our case
that we do not see any absorption lines caused by the predominantly red stars
of the clusters or by moon light and that the night sky emission lines in
the red part could be corrected down to a level of a few percent.
The fitted sky background was then subtracted from the unsmoothed frame
and the spectra were extracted using Horne's algorithm (Horne, \cite{horn86})
as implemented in MIDAS.

Finally the spectra were corrected for atmospheric extinction and response
curves were derived from the spectra of the flux standard stars. Normally
the response curves were fitted by a 3$^{\rm rd}$ order spline, except
the red EMMI data, for which we used a 6$^{\rm th}$ order polynomial.
We took special care to correctly fit the response curves for the blue
low resolution data in the region of the Balmer jump, since we will use this
feature for the determination of the effective temperatures of the stars.

Comparing the EFOSC1 and EMMI low resolution data for the stars observed
during both runs showed that the EFOSC1 data have significantly less
UV flux blueward of the Balmer jump than the EMMI data. Since
both sets of spectra could be fitted with Kurucz (\cite{kuru92}) models
(but yielded temperatures different by up to 10000 K)
we could not decide which data set was the correct one and whether
one or even both data sets were hampered by instrumental effects
that were not removed by the described reduction. We discussed these
problems extensively with people at ESO and finally decided to repeat the
spectrophotometric observations with a third instrument (which decision
resulted in the 1995 observations with EFOSC2).

The medium resolution spectra observed in 1992 and 1993 are plotted in
Fig.~2 and 3. Besides the Balmer lines, \ion{He}{I} lines can easily be
identified in the spectra of several of the programme stars. In addition
\ion{He}{II}, 4686~\AA , is visible in B~852, which therefore has to be
classified as spectral type sdOB according to the scheme of Baschek \&
Norris (\cite{bano75}). It is also worthwhile to note that weak \ion{Mg}{II},
4481~\AA , absorption is detected in B~1509 and B~3975

We also used the medium resolution data to derive radial velocities,
which are listed in Table~2 (corrected to heliocentric system).
The error of the velocities is about
30~km/sec. Within error limits the radial velocities agree with the cluster
velocity ($-$32.1 km/s, Pryor \& Meylan, \cite{prme93}).

\subsection{EFOSC2 data}

For the EFOSC2 data we averaged the bias and flat field frames over both nights
of the run, as they showed no deviations above the 1\% level.
The bias correction was again
performed by scaling the bias with the mean overscan of the frame and we did
not correct for dark current. The normalization of the flat-field was
somewhat difficult: We fitted a 7$^{\rm th}$ order polynomial to the flux
distribution of the averaged flat fields, which did not fit well enough
the UV/blue part. We therefore averaged and smoothed the flat as described
for the EFOSC1 data and merged the UV/blue part of the smoothed averaged flat
and the red part of the fit at a position where the two overlapped and had a
similar slope. This merged
frame was smoothed again before correcting the flat.

For the wavelength calibration we again used
a 3$^{\rm rd}$ order polynomial to fit the dispersion relation. However,
while the wavelength calibration frames showed maximum deviations of
4~\AA\ when rebinned using this dispersion relation, the science frames
exhibited systematic shifts to the red in the order of 30 -- 40~\AA.
The only difference between the two types of frames (beside the illumination)
was the slit width, which was 1\arcsec\ for the calibration
frames and 5\arcsec\ for the science frames.
A possible explanation of this effect could lie with the fact that the
polynomial fits normally used for wavelength calibrations are only an
approximation to the true shape of the dispersion relation, which is
described correctly by trigonometric functions (Strocke \cite{stro67};
Bahner \cite{bahn67}). Tests performed by
Rosa \& Hopp (1995, priv. comm.) showed, however, that fits using such
trigonometric functions are extremely unstable unless a very good
guess of the true relation is already available.
We therefore decided to correct the offsets by applying a mean shift of
36~\AA\ to the blue.

Sky subtraction, correction of atmospheric extinction,
 and flux calibration were performed in the same way as for
the EMMI and EFOSC1 data, using a constant for the spatial profile of the
sky background and a spline fit for the response curve.
As mentioned above we could judge the photometric quality of the
observations from the response curves. We used only those curves that yielded
the maximum response
since those should result from observations taken without any obscurations.
Several object spectra calibrated with these response curves
showed an offset from the B and V magnitudes
taken from Buonnano et al. (\cite{buca86}), being about 0.2 mag too faint.
Although there was hardly any wavelength dependency of this offset
visible we could not be sure that the absorption was truly grey.
We therefore decided not to use these data for determinations of the effective
temperatures because we could not rely on the
continuum slope. They are marked in Table~2 by $\otimes$.

\subsection{Comparison of the optical spectrophotometric data}

Comparing the data for the three runs shows that the EMMI and the
EFOSC2 data agree rather well when a correction for the much lower
resolution of the EFOSC2 data is applied (necessary to allow comparison
of the Balmer jump region). This leads to the assumption that there is
some problem with the UV response of the EFOSC1 that is not corrected
by flux calibration. An additional reason could lie with the extraordinarily
high extinction observed at La Silla in the years 1991 and 1992 (Burki et
al. \cite{buru95}). We therefore ignore all low resolution data
obtained with EFOSC1 for the further analysis.
As a consequence, 5 (B~210, B~852, B~1509, B~1628, B~4380)
out of 17 programme
stars lack reliable spectrophotometric data. For an additional star (B~3915)
we do not have any low resolution data at all.

\section{Atmospheric Parameters}

Effective temperatures, gravities, and helium abundances are derived from
line blanketed LTE model atmospheres. { We used the
Kurucz (\cite{kuru92}) grid and models calculated with
an updated version of the code of Heber (\cite{hebe83}) using ATLAS6 opacity
distribution functions (ODF),
log(Z/\Zsolar)~=~$-$2.0 and helium abundance of 0.1 solar, appropriate
for blue horizontal branch stars . }

Kurucz (\cite{kuru92}) ATLAS9 models (log\,g~=~5.0, solar He abundance,
log(Z/\Zsolar)~=~$-$1.5 closest to the cluster's metallicity, [Fe/H]~=~$-$1.54,
Djorgovski, 1993) were used to fit the energy distributions (Sect. 4.1)
because of their superior spectral resolution in the region of the Balmer
jump.
To correct for interstellar
reddening we applied the reddening law of Savage \& Mathis (\cite{sama79})
and used an E$_{\rm B-V}$ of 0.05 mag.

Heber's models were used to analyse the hydrogen and helium line spectrum
(Sect. 4.2 and 4.3) in order to account for the appropriate gravities and
helium abundances.
{
As these models  were computed for [M/H] = $-$2.0
(instead of $-$1.5) and sub-solar helium abundance
 we fitted the Kurucz hydrogen line profiles
(\teff\ =~22000~K, \logg\ =~5.0) by Heber models of log(Z/\Zsolar)~=~$-$2.0.
The use of different generations of ODFs (ATLAS9 versus
ATLAS6) and different helium abundances
turns out to be negligibly small, both for \teff\ and \logg .
Hence, the following procedure is applied: The Balmer lines
H$_\beta$, H$_\gamma$, and H$_\delta$ are fitted using Heber's models.
At the final parameters (\teff,
\logg)  curves of growths for the \ion{He}{I} lines
4026, 4121, 4388, 4471 and 4921~\AA\ are constructed to determine the helium
abundances from equivalent widths of these lines.

In order to allow a consistent treatment of the results of Heber et al. (1986),
which were obtained with solar metallicity models, we checked the effect of
the metallicity on \teff\ and \logg\ by fitting solar metallicity models
with the [M/H] = $-$2.0 models we use now. We found out that the solar
metallicity models yield effective temperatures about 1000~K lower than
the metal poor models, but essentially the same \logg\ values. We therefore
increased the \teff\ values of Heber et al. (1986) by an offset of 1000~K to
account for the lower cluster metallicity and kept the \logg\ values.
}

In the case of B~852 LTE models were found to be inappropriate due to its
high \teff\ (see Napiwotzki, \cite{napi96}). We used his NLTE models which
were calculated with the code of Werner \& Dreizler (\cite{WeDr96}) and
do not
include metal line blanketing. Therefore the \teff\ scale slightly differs
from the metal line blanketed scale of Kurucz and Heber models. No attempt
was made for adjustment of \teff\ scales.

\begin{figure*}
\includegraphics{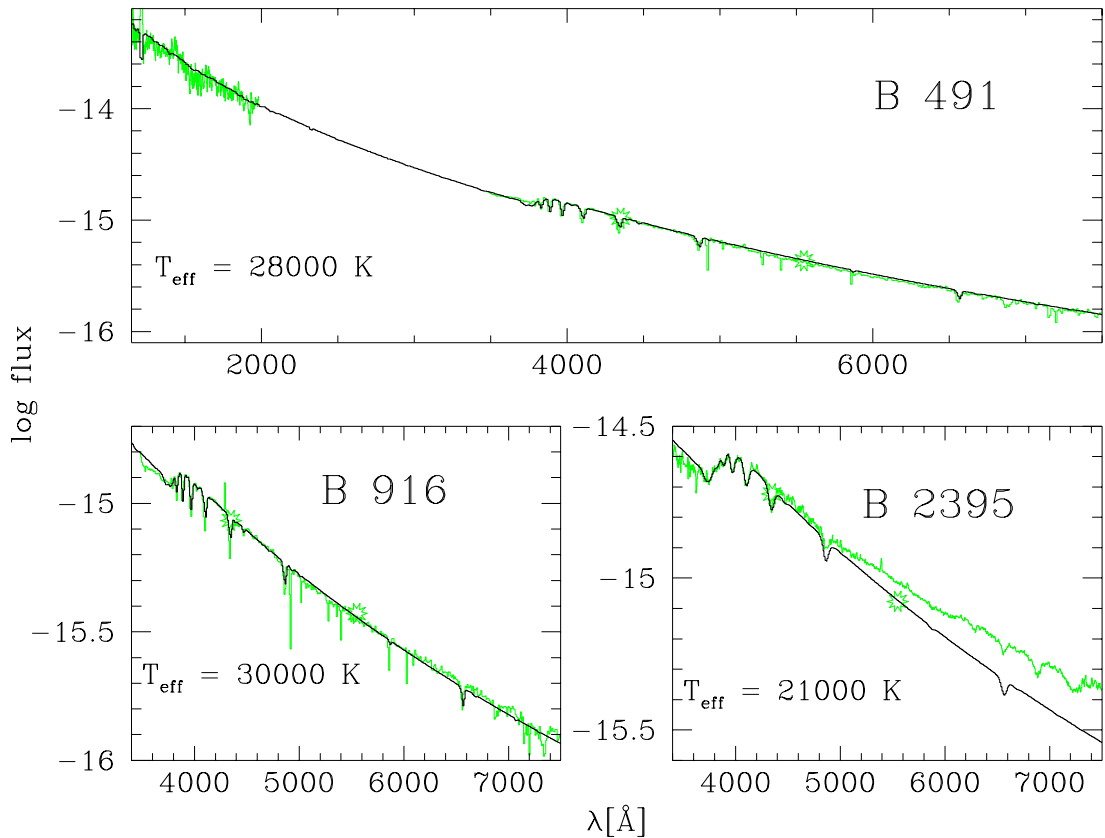}
\vspace{12cm}
\caption[]{Fits to low resolution data of B~491, B~916, and B~2395. The models
are ATLAS9 (Kurucz, 1992) models of metallicity $-$1.5, all for \logg\ =~5.0.
{ The asterisks mark the B and V fluxes derived from Buonanno et al. (1986)
using the conversion factors given by Heber et al. (1984). The fluxes as well
as
the observed spectra are dereddened using the reddening law of Savage \&
Mathis (1979).}}
\end{figure*}

\subsection{\teff\ derived from energy distributions}

\subsubsection{IUE data and the results of Cacciari et al. (1995)}

For the four stars which have IUE spectra we tried to simultaneously fit
the continuum data (IUE spectra, BV photometry, continuum of
the low resolution optical spectra) and the Balmer jump (cf. Fig 4).
In two cases the Balmer jump suggested
temperatures different from those derived from the continuum data { (B~491
and B~916, cf. Table~3).

The same data have already been analysed by Cacciari et al. (1995) by
calibrating UV- (1300\AA\ -1800\AA) and UV-visual (1800\AA\ - V) colors in
terms of \teff\ and \logg\ using Kurucz model atmospheres.
In all cases our effective temperatures are lower
than the ones given by Cacciari et al. (\cite{cafu95}).
This can be traced back to the use of different IUE flux calibrations. We used
the most recent calibration of Bohlin (1996) which gives a flux at 1800\AA\
lower by 10\%\ than one of
the Cacciari et al. (1995) and, hence, changing the (18-V)
colour by 10\% , which translates in a \teff\ decrease exactly as found when
comparing our results to those of Cacciari et al.

In view of this flux calibration problem, we also went back to the IUE data
for the stars analysed by Heber et al. (1986) and fitted the recalibrated
IUE spectra and the optical photometric
data as described above.
The new temperatures were
always within 1000~K of the old values and no systematic trend showed up.
We therefore decided to keep the old values and only adjust them for
metallicity effects (see above).

}

\subsubsection{Optical data only}

The simultaneous fitting of the spectrophotometric data (Balmer jump and
continuum) and the B and V magnitudes worked well (cf. Fig.~4)
in all cases except four: { B~1628, B~2395, B~3655, and B~2126.
B~1628 shows a strong red excess when compared to the B and V magnitudes and
to the model fitting of the Balmer jump; the observed spectrum is also
significantly brighter than the B and V magnitudes.
The stars B~2395 (cf. Fig.~4)
and B~3655 both show a moderate red excess when compared to
the model fitting the Balmer jump, which also reproduces the B and V
magnitudes.
In the case of B~2126 the optical spectrum agrees well with BV photometry, but
both show a small red excess when compared to the model fitting
the Balmer jump.

Cautioned by these findings we searched all stars for red neighbours.
For this purpose we used the photometry of NGC 6752 by Buonanno et al.
(\cite{buca86}), and
extracted for each of our targets all stars within a radius
of 20\arcsec (down to the limiting magnitude
of about B~$\approx$~\magpt{18}{5}).
We then estimated the straylight provided by these stars
assuming that the seeing at larger distances is
best described by a Lorentz profile and scaling the intensities with the V
fluxes. As seeing value we used 1.5\arcsec as ``intrinsic seeing''
for all observations (which
overestimates the seeing near the meridian) and took
into account the elongation caused by atmospheric dispersion.
Under these assumptions we get stray light levels of more than 3\% in
V for the following objects: B~617, B~1288, B~1628,
B~2162, B~2395, and B~3655.
As expected we found the largest values for B~1628, B~2395, and B~3655,
which were observed at relatively large zenith distances and have very bright
neighbours.
For those objects where the neighbour did not lie in the slit the observed
stray light levels are lower than the calculated ones by a factor of 1.5 to 2.
The calculated
excesses for B~1288 and B~617 are about 7\% and cannot be measured
in our spectrophotometric data.
Hence we can explain the above mentioned red excesses as being caused by
stray light from nearby stars (except B~2126). At the same time we can derive
an upper limit of 7\%\
for the accuracy of our spectrophotometric calibration (at V).

Since the Balmer lines (H$_\beta$, H$_\gamma$, H$_\delta$) are diagnostic
tool we also checked the stray light level in the medium resolved spectra.
We used the
same assumptions as above, but scaled the intensities with the B magnitudes.
This results in
straylight levels of at most 5\%\ for B~617, B~1288,
B~1509, B~1628, B~2162, B~2395, and B~3655.
In order to estimate the effect of any additional
continuum on the physical parameters we subtracted from the spectrum of B~2395
a constant offset of 5\%, renormalized the spectrum and derived \teff\
and \logg\ again from only fitting the Balmer line profiles. It turned out
that the best fit yielded an effective temperature of 1000~K less
than before and the surface gravity remained the same.
}

\begin{figure}
\includegraphics{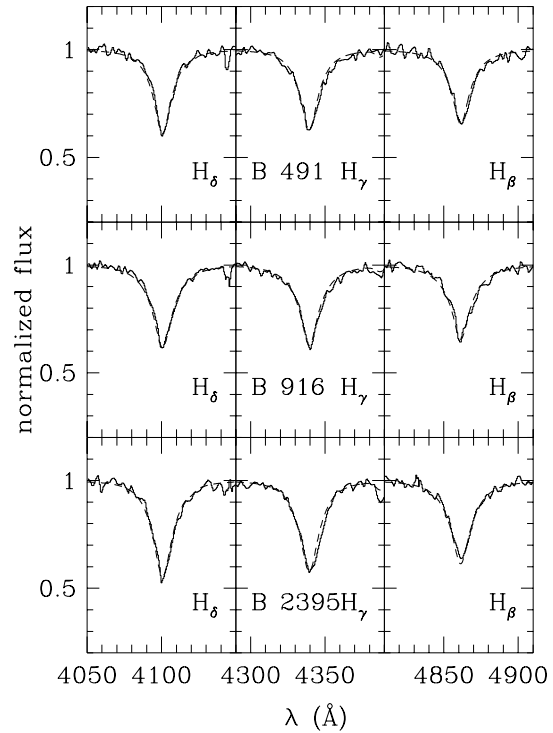}
\vspace{9.5cm}
\caption[]{Fits to medium resolution data of B~491, B~916, and B~2395.
The models
are those of Heber (1983) for those parameters which best fitted the Balmer
lines (\teff\ is already transformed to the Kurucz scale):\\
\begin{tabular}{ll}
B~491: & \teff\ =~28000~K, \logg\ =~5.2\\
B~916: & \teff\ =~30000~K, \logg\ =~5.6\\
B~2395: & \teff\ =~22000~K, \logg\ =~5.0\\
\end{tabular}}
\end{figure}

\begin{figure}
\includegraphics{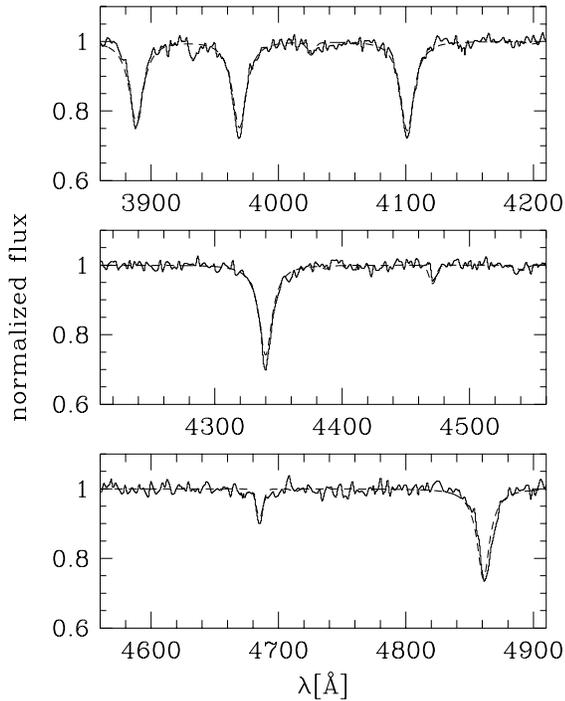}
\vspace{9.5cm}
\caption[]{Fit to medium resolution data of the sdOB star B~852.
Overplotted is a Non-LTE model
computed by R. Napiwotzki (priv. comm.) for a metal-free stellar atmosphere
with a 0.1 solar helium abundance.}
\end{figure}

\begin{table*}
\begin{tabular}{|l|rr|rrr|rrrr|}
 \hline
Star & V & B$-$V & \teff $_{,UV}$ & \teff $_{,opt.}$ & \teff $_{,lines}$
& \teff $_{fin}$ & \logg\ & M  & M$_V$ \\
 & [mag] & [mag] & [K] & [K] & [K] & [K] & & [\Msolar ] & [mag] \\
 \hline
\multicolumn{10}{|c|}{Blue HB stars}\\
\hline
B~577$^1$ & 14.85 & $-$0.04 &     &       &       & 13700 & 3.9 & 0.43 & 1.61\\
B~1509     & 15.52 & $-$0.06 &     &       & 17000 & 17000 & 4.1 & 0.26 &
2.28\\
B~2454$^1$ & 14.27 & $-$0.06 &     &       &       & 10700 & 3.5 & 0.45 &
1.03\\
B~4104$^1$ & 15.20 &   +0.00 &     &       &       & 17000 & 4.0 & 0.27 &
1.96\\
B~4719$^1$ & 15.65 & $-$0.06 &     &       &       & 16000 & 4.0 & 0.20 &
2.41\\
\hline
\multicolumn{10}{|c|}{Stars in the gap region}\\
\hline
B~1628 & 16.30 & $-$0.17 &       &       & 21000 & 21000 & 4.7 & 0.35 & 3.06\\
B~2395 & 16.73 & $-$0.24 &       & 21000 & 22000 & 21500 & 5.0 & 0.46 & 3.49\\
B~3655 & 16.40 & $-$0.22 &       & 24000 & 22000 & 23000 & 5.1 & 0.70 & 3.16\\
B~3975 & 16.27 & $-$0.22 &       & 21000 & 21000 & 21000 & 4.8 & 0.46 & 3.03\\
B~4548 & 16.60 & $-$0.14 & 22000 & 23000 & 23000 & 22500 & 5.2 & 0.75 & 3.36\\
\hline
\multicolumn{10}{|c|}{EHB stars}\\
\hline
B~210    & 17.04 & $-$0.31 &       &       & 27000 & 27000 & 5.6 & 0.94 &
3.80\\
B~331$^1$ & 17.12 & $-$0.22 &      &       &       & 26000 & 5.6 & 0.93 &
3.88\\
B~491    & 17.45 & $-$0.31 & 28000 & 30000 & 28000 & 28500 & 5.3 & 0.29 &
4.21\\
B~617$^2$ & 17.84 & $-$0.33 &      & 33000 & 26000 & 33000 & 6.1 & 0.98 &
4.60\\
B~763$^1$ & 17.13 & $-$0.24 &      &       &       & 27000 & 5.5 & 0.68 &
3.89\\
B~916    & 17.61 & $-$0.26 & 27000 & 30000 & 30000 & 28500 & 5.4 & 0.32 &
4.37\\
B~1288   & 17.90 & $-$0.31 &       & 29000 & 27000 & 28000 & 5.5 & 0.31 &
4.66\\
B~2126   & 17.28 & $-$0.20 &       & 31000 & 28000 & 29500 & 5.7 & 0.79 &
4.04\\
B~2162   & 17.88 & $-$0.27 &       & 35000 & 32000 & 33500 & 5.9 & 0.57 &
4.64\\
B~3915   & 17.16 & $-$0.23 &       &       & 31000 & 31000 & 5.5 & 0.47 &
3.92\\
B~4009   & 17.44 & $-$0.30 & 33000 & 29000 & 31000 & 31500 & 5.7 & 0.61 &
4.20\\
3-118$^1$& 17.76 & $-$0.26 &       &       &       & 24500 & 5.2 & 0.23 &
4.52\\
\hline
\multicolumn{10}{|c|}{Post-EHB stars}\\
\hline
B~852 & 15.91 & $-$0.28 &      &       & & 39000 & 5.2 & 0.53 & 2.67\\
B~1754$^1$ & 15.99 & $-$0.24 &     &       &       & 40000 & 5.0 & 0.29 &
2.75\\
B~4380     & 15.93 & $-$0.14 &     &       & 32000 & 32000 & 5.3 & 0.93 &
2.69\\
\hline
\end{tabular}
\caption{Physical parameters of the observed stars. The star numbers
{ and V, B$-$V data are taken from
Buonanno et al. (1986), except for 3-118, whose data were obtained from Caloi
et al. (1986).}
\newline
$^1$: Data from Heber et al. (1986), adjusted to the Kurucz temperature scale.
{
Note that the mass value for B~1754 given by Cacciari et al. (1995) results
from a incorrect bolometric correction (Cacciari, priv. comm.)
\newline
$^2$: Since B~617 lies close to B~1288 and B~2162 in the CMD we believe that
the higher temperature is the correct one (see text). It was not taken into
account for the mass distribution.
}
}
\end{table*}

\subsection{Balmer line profile fits}

To check the validity of the effective temperatures obtained from the
low resolution data we also determined \teff\ and \logg\
simultaneously by fitting the line profiles of H$_\beta$, H$_\gamma$, and
H$_\delta$.

In most cases where low and intermediate resolution spectra are available the
effective temperatures derived from both agree within the error limits
(cf. Fig.~5 and Table~3).
In one case (B~617) however, we find an effective temperature from
the low resolution data of about 33000~K, while the line profiles
yield 25000~K. Since this star is in colour and brightness comparable to
B~1288 and B~2162, we assume that the higher temperature is the valid
one.
The systematic offset between \teff\ derived from low resolution data and
from the Balmer line profiles reported in paper~I
could not be found with these data.

{ In order to get an estimate for the internal errors of the physical
parameters derived from fitting the Balmer lines we used the
independent line fitting code of R. Saffer
(see Saffer et al., \cite{saff94}) to verify our results.
It turned out that we got the same values for spectra with good S/N. For
the spectra of 1992, which are considerably more noisy than the ones from 1993,
we get systematically lower temperatures with Saffer's routine than from
our line profile fits (using the same set of models). Since we cannot reconcile
the spectrophotometric data with the temperatures found by Saffer's routine we
decided to keep our values. The mean internal
error in \teff\ provided by Saffer's
fits are 1370 K (1992) and 460 K (1993); the mean internal error in \logg\
is 0.17 dex (1992) and 0.06 dex (1993).
}

\subsection{The case of B~852}

B~852 is the only star in our sample which shows \ion{He}{II}, 4686~\AA.
Comparing its strength to that of \ion{He}{I}, 4471~\AA\ suggests a
temperature exceeding 35000~K. At such high temperature deviations from LTE
become important even for the relatively large gravities of hot subdwarf
stars as demonstrated by Napiwotzki (1996). Therefore we used his NLTE
models to derive \teff, \logg, and helium abundance from
H$_\beta$, H$_\gamma$, and H$_\delta$,  \ion{He}{II}, 4686~\AA\ and
\ion{He}{I}, 4471~\AA, the latter line ratio being a very sensitive \teff\
indicator. The best fit  is shown in Fig. 6. The depth of
the observed Balmer line profile cores cannot be reproduced by the NLTE
models, which we attribute to the lack of metal line blanketing.
The helium lines are not affected by this effect and are reproduced
very well. We therefore use the best fit values
of \teff\ =~40000~K, \logg\ =~5.2
and log(n$_{\rm He}$/n$_{\rm H}$)~=~$-$2.0 as atmospheric parameters for B~852.

\subsection{Final parameters}

As effective temperature we finally used the mean value of all available
determinations, assessing the IUE temperatures double weight
{ because of the superior sensitivity to \teff\ of UV data.
Besides the internal errors quoted above  we also have to account
for systematic errors caused by the imperfections of models and data. These
are estimated  from the differences between
various \teff\ determinations for the same object to be about 7\% in
\teff\ (cf. Table 3).}

For the given final temperature we take that value of \logg\ as
surface gravity that yields the smallest errors for the Balmer line fits at
this temperature. { An error in \teff\ of 7\%\ translates into an error
in \logg , which we estimated for the three groups of stars mentioned in
Table 3:  0.1 dex for BHB stars ($<$\teff $>$ = 15000~K); 0.15~dex for
stars in the gap region ($ <$\teff $>$ = 22000~K) and
0.2 dex for EHB stars ($<$ \teff $>$ = 29000 K).
Added to these errors in \logg\ are the internal errors given by Saffer's
fitting routine (see Sect. 4.2). For the data given by Heber et al.
(\cite{heku86}) we used the mean of the errors for the 1992 and 1993 data.
This results in the following errors in \logg : BHB stars (1992: 0.196;
1993: 0.115); stars in the gap region (1992: 0.225; 1993: 0.160);
EHB stars (1992: 0.261; 1993: 0.208).
}
Absolute visual magnitudes were derived from the apparent
magnitude and the distance modulus
((m$-$M)$_{\rm V}$~=~\magpt{13}{24}, Djorgovski \cite{djor93}).
{ Note that th reddening free distance modulus of \magpt{13}{12} of
Djorgovski agrees within the error limits with the new distance modulus
(m$-$M)$_{\rm V,0}$~=~\magpt{13}{05} derived by Renzini et al. (\cite{rebr96})
by fitting the white dwarf sequence. }
The results are listed in Table~3.

\subsection{Helium abundances}

{}From the medium resolution spectra we could also determine the helium
abundances
of the stars (in case helium lines are seen) or at least upper limits.
The measured equivalent widths of the \ion{He}{I} lines are given in
Table~4. Only one star (B~852) shows also \ion{He}{II} 4686~\AA .
Using the physical
parameters listed in Table~3 we then derive helium abundances. The
spectra observed in 1992 were rather noisy and therefore equivalent widths of
less than 0.35~\AA\ cannot be measured. The data taken in 1993 are of much
better quality and allow equivalent widths as low as 0.15~\AA\ to be
measured.
We take these values (0.35~\AA\ resp. 0.15~\AA ) as upper limits if we cannot
detect a helium line in a spectrum and derive an upper limit for the helium
abundance from the absence of \ion{He}{I}, 4471~\AA, the helium line
predicted to be strongest in the observed wavelength range.
This upper limit is determined for each object individually.
All stars are
helium deficient with respect to the sun by factors ranging from 3 to more
than 100 and no trends with \teff\ or \logg\ are
apparent, similarly to the field sdBs (as demonstrated
by Schulz et al., \cite{schu91}).

\begin{table*}
\begin{tabular}{|l|l|l|l|l|l|l|l|r|r|}
\hline
Name & year & 4026 \AA & 4120 \AA & 4144 \AA & 4388 \AA & 4471 \AA & 4921 \AA
& log n$_{\rm He}$ & n$_{{\rm He}, \odot}/$n$_{\rm He}$ \\
 & & [\AA ] & [\AA ] & [\AA ] & [\AA ] & [\AA ] & [\AA ] & &\\
\hline
B 210 & 1992  & 0.75 & -    & -    & -    & 1.06 & - & $-$2.0    & 10 \\
B 491 & 1992  & -    & -    & -    & -    & -    & - & $<-$3.2 & $>$160 \\
B 617 & 1992  & -    & -    & -    & -    & -    & - & $<-$2.4 & 25 \\
B 852$^1$ & 1993 & - & -    & -    & -    & 0.26 & - & $-$2.0    & 10 \\
B 916 & 1993  & 0.90 & 0.07 & 0.30 & -    & 1.40 & 0.40  & $-$1.8   &  4 \\
B 1288 & 1992 & -    & -    & -    & -    & 1.57 & 1.00 & $-$1.5 &  3 \\
B 1509 & 1992 & 0.34:& -    & -    & -    & -    & - &  & \\
$^2$   & 1993 & 0.46 & -    & 0.12*& -    & 0.45 & 0.18* & $-$2.2&  15\\
B 1628 & 1993 & 0.28 & 0.14 & 0.26 & 0.22*& 0.38 & 0.31 & $-$2.4 &  25\\
B 2126 & 1992 & -    & -    & -    & -    & -    & - & $<-$3.0 & $>$100\\
B 2162 & 1993 & 0.49 & -    & -    & -    & 0.80 & 0.45 & $-$1.8 &    6 \\
B 2395 & 1993 & 0.79 & -    & -    & 0.73 & 0.90 & 0.35 & $-$1.9 &    7 \\
B 3655 & 1992 & 0.80 & -    & -    & -    & 0.74 & 0.95 & $-$1.8 &    7\\
B 3915 & 1993 & -    & -    & -    & -    & -    & - & $<-$3.0  &$>$100\\
B 3975$^3$ & 1993 & 0.73 & -    & 0.22 & 0.27 & 0.94 & 0.19* & $-$2.2  &  15\\
B 4009 & 1993 & -    & -    & -    & -    & -    & - & $<-$3.1 & $>$ 120\\
B 4380 & 1992 & -    & -    & -    & -    & 0.38:& - & $-$2.3  & 20 \\
B 4548 & 1993 & 0.87 & 0.36:& 0.97:& -    & 0.95 & - & $-$1.9  &  8\\
\hline
\end{tabular}
\caption[]{The equivalent widths of the \ion{He}{I} lines as measured from
the medium resolution spectra and the corresponding helium abundances
(in particle numbers).\\
$^1$ B~852 shows also \ion{He}{II} at 4686~\AA\ (W$_\lambda$~=~0.8~\AA)\\
$^2$ B~1509 (1993) shows also \ion{Mg}{II} 4481 ~AA\ (W$_\lambda$~=~0.23~\AA
)\\
$^3$ B~3975 shows also \ion{Mg}{II} 4481~\AA\ (W$_\lambda$~=~0.26~\AA )
and \ion{He}{I} 4169~\AA (W$_\lambda$~=~0.13:~\AA )\\
: equivalent width has a large error due to a rather noisy spectrum\\
$*$ equivalent width is at the level of the measuring error}
\end{table*}

\subsection{Evolutionary status}
The results are compared in Fig.~7 (\teff, \logg)
and 8 (\teff, M$_{\rm V}$)
to the predictions of stellar evolution theory (Dorman et al. 1993).

The bulk of the stars hotter than 20000 K lies within the
theoretically predicted region (within the error bars), except three
(B~852, B~1754, and B~4380).
The latter three stars have about the same magnitude
(V~=~\magpt{15}{91}--\magpt{15}{99})
and lie above the gap in the CMD, albeit far to the blue edge of the HB in this
part. They are probably in a Post-EHB stage of evolution (cf. Fig.~7), as
already argued
by Heber et al. (\cite{heku86}) in the case of B~1754, and will evolve
directly towards the white dwarf cooling sequence.
The distribution of stars in the (\teff, M$_V$) diagram (Fig. 8) is
consistent with that in the (\teff, \logg)-plane (Fig. 7) indicating that
the stars above 20000~K (except the three mentioned above) are
bona fide EHB stars.
The twelve stars below the gap (V~$>$~\magpt{17}{0}) have a mean absolute
magnitude of M$_{\rm V}$~=~\magpt{4}{2}~$\pm$~\magpt{0}{3}
in excellent agreement with the mean absolute magnitude of
south galactic pole sdBs, M$_{\rm V}$~=~\magpt{4}{2} $\pm$ \magpt{0}{7}
(Heber, \cite{hebe86}).
The five stars in the gap
region (\magpt{16}{0}~$<$~V~$<$~\magpt{17}{0}) have \teff\ $>$~20000~K, \logg\
near 5 and subsolar helium abundances. Therefore they show the same
characteristics as the field sdB stars and have to be
classified as sdBs too. However, they are somewhat more luminous
(M$_{\rm V}$~=~\magpt{3}{2}~$\pm$~\magpt{0}{2})
than the stars below the gap in NGC 6752 and the field sdBs.

The fact that the stars between 10000~K and
 20000~K lie systematically above the HB region in the (\teff, \logg)
diagram has already been noted
in papers~I and II and also by Crocker et al.
(\cite{crro88}).

\begin{figure}
\includegraphics{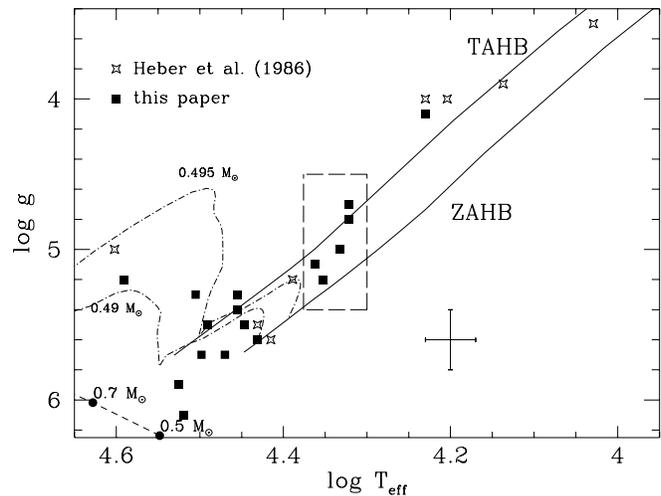}
\vspace{6.8cm}
\caption[]{The resulting physical parameters listed in Table~3
compared to theoretical predictions. The solid lines are the Zero Age HB and
the
Terminal Age (core helium exhaustion) HB for [Fe/H]~=~$-$1.48 of Dorman
et al. (1993). The long dashed line marks the gap region seen in the
CMD by Buonanno et al. (1986). The short dashed line gives the position of the
helium main sequence (Paczynski 1971). The long dashed-short dahed lines
give post-EHB evolutionary tracks by Dorman et al. (1993), labeled with
the total mass of the EHB star.}
\end{figure}

\begin{figure}
\includegraphics{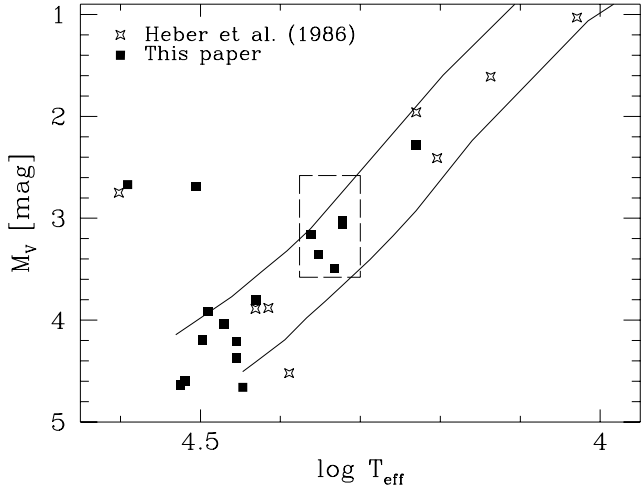}
\vspace{6.8cm}
\caption[]{The absolute V magnitudes and effective temperatures
listed in Table~3
compared to theoretical tracks by
Dorman et al. (1993, details see Fig.~7).
The long dashed line marks the gap region seen in the
CMD by Buonanno et al. (1986).}
\end{figure}

\section{Masses}

\begin{figure}
\includegraphics{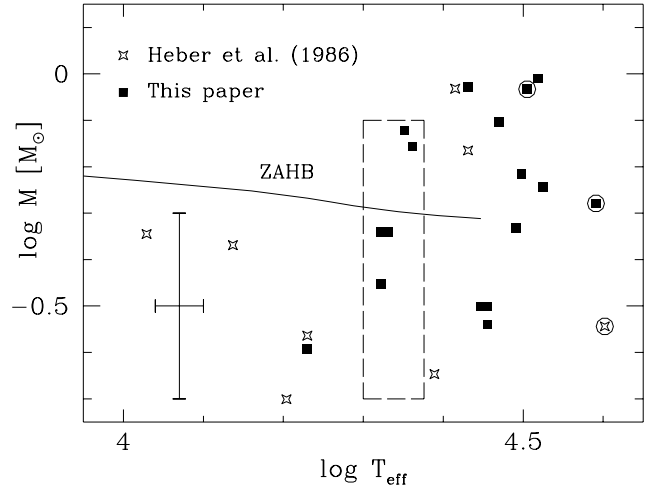}
\vspace{6.3cm}
\caption[]{The resulting logarithmic masses listed in Table~3
vs. \logt . The objects that were
not used to derive the mean sdB mass (B~617, B~852, B~1754, B~4380) are
marked by circles. Also plotted is
the Zero Age HB for [Fe/H]~=~$-$1.48 of Dorman et al. (1993)
The long dashed line marks the gap region seen in the
CMD by Buonanno et al. (1986).}
\end{figure}

Knowing \teff , \logg , and the distances of the stars we can derive the
masses as described in paper~I:

\noindent
{
$\log\ (M/\Msolar) = const. + \log\ g + 0.4 \cdot ((m-M)_V -V + V_{th})$

\noindent
(V$_{th}$ denotates the theoretical brightness at the stellar surface
as given by Kurucz \cite{kuru92}.)}

The
results are listed in Table~3 and plotted in Fig.~9.
{ In addition to the error in \logg\ errors in the absolute magnitude
and the theoretical brightness at the stellar surface also enter the final
error in log~M. We assume an error in the absolute brightness of \magpt{0}{1}
for stars with V~$\le$~\magpt{17}{0} and \magpt{0}{13} for fainter stars.
The error
in the theoretical V brightness is dominated by the errors in \teff\
and amounts to \magpt{0}{12} (BHB stars); \magpt{0}{13}
(stars in the gap
region); and \magpt{0}{14} (EHB stars).
Alltogether this leads to the following errors in log~M: BHB stars (1992:
0.210; 1993: 0.137); stars in the gap region (1992: 0.238; 1993:
0.178); EHB stars (1992: 0.275; 1993: 0.225).}
To determine the mean mass of the sdBs
(this paper and Heber et al., \cite{heku86}) we took all stars with
\teff\ $>$~20000~K, excluding B~617 (due to the strange offset between
temperatures from continuum and from line profiles)
and the post-EHB stars B~852, B~1754, B~4380,
resulting in a total of 16 stars.
{
We then calculated the weighted mean of the logarithmic masses for these stars
(the weights being derived from the inverse errors).
The mean logarithmic mass for these stars then is -0.300 (= 0.500~\Msolar).
The logarithmic standard deviation
is 0.043 dex and the expected mean logarithmic error
as derived from the observational errors is 0.054 dex.
The mean mass therefore agrees extremely well
with the value of 0.488~\Msolar\ predicted by canonical HB
theory for these stars (Dorman et al., 1993)
and the standard deviation is less than expected.
If we omit the stars observed in 1992 (due to their higher errors)
we get a mean logarithmic mass for the remaining 12 stars
of -0.327 (= 0.47~\Msolar), which is
somewhat lower than the result above but still in very good agreement with
theoretical predictions.
There is no significant difference between the mean mass for the eleven
stars below the gap ($<$log~M$>$ = -0.306; $<$M$>$ = 0.494~\Msolar)
 and the five stars
inside the gap region ($<$log~M$>$ = -0.289; $<$M$>$ = 0.514~\Msolar);
again pointing towards their nature being
identical.
The five stars above the gap (V~$<$~\magpt{16}{0}, \teff\ $<$~20000~K)
have a mean logarithmic mass of
$-$0.518 (= 0.303~\Msolar)
with a logarithmic standard deviation of 0.059 (compared to an expected
error of 0.073).
Canonical theory would predict for these stars a mean logarithmic mass of
$-$0.256 (=~0.554~\Msolar) with a scatter of 0.018 dex.
}
\section{Conclusions}

Combining our results with those of Heber et al. (\cite{heku86}) we have
carried out spectroscopic analyses of 25 blue stars in NGC 6752. We studied
stars both brighter and fainter then the gap region
(\magpt{16}{0}~$<$~V~$<$~\magpt{17}{0}) and also five stars
within this gap region. The
results can be summarized as follows:
\begin{itemize}
\item[(i)] All stars are helium deficient and no trends of the helium
abundance with atmospheric parameters become apparent.
\item[(ii)] All stars below the gap (except three) lie on or close
to the Horizontal Branch.
The latter three stars have already evolved off the EHB towards the white
dwarf cooling sequence.
\item[(iii)] The stars fainter than the gap (V $>$ \magpt{17}{0}) are
very similar to the field population of sdB stars. Their mean mass is
consistent with the canonical value of half a solar mass.
\item[(iv)] The stars inside the gap region
(\magpt{16}{0}~$<$~V~$<$~\magpt{17}{0}) are sdB stars too, although more
luminous by one magnitude than the average field sdBs. Their mean mass is
also consistent with the canonical mass. Their position in the
(\teff, \logg)-plane
agrees less well with HB evolutionary calculations of Dorman et al. (1993)
than the fainter EHB stars do.
\item[(v)] The stars brighter than the HB gap have significantly lower
gravities than predicted by theory. Their masses are also significantly
lower than predicted by evolutionary calculations.
{ \item[(vi)] The standard deviations
of the mean logarithmic masses for all groups
of stars are lower than expected from the observational errors.}
\end{itemize}

{}From these results we conclude:

\begin{itemize}
\item[(i)] The helium deficiency found in all stars is indicative for
diffusion, i.e. gravitational settling of helium as is the case for the
field sdB and BHB stars.
\item[(ii)] The increased sample size allowed a more meaningfull
determination of the mean mass of the
sdB stars than pevious work did and it turned out to be in good agreement
with the
value predicted by classical EHB theory. { In addition, the standard
deviation is
smaller than the mean error derived from the observational errors.}
We take this as a verification that sdB
stars { indeed show a very narrow mass distribution, which is in full
agreement with their status as }
extreme HB stars with a very thin hydrogen layer, produced
by the same processes that produce BHB stars.
This finding, however, does not explain the gap that
separates BHB from sdB stars in NGC~6752 (and also
in M~15, Durrell \& Harris \cite{duha93}).
The mean mass and the width of the mass distribution
is consistent with the single star evolutionary scenario or
the Mengel et al. binary evolution scenario (No. II in the introduction)
for the origin of EHB stars,
but can hardly be explained within the framework of the binary scenario (I)
and the merger scenario (III) listed in the introduction, since both
scenarios would predict a broader mass
distribution with a  mean mass below 0.5\,\Msolar.
\item[(iii)] The low masses { as well
as the low surface gravities} we find for the BHB stars (brighter than the gap)
remain a puzzle.
{
Lower than expected gravities for BHB stars were also found recently by Dixon
et al. (\cite{wvd96}) in the globular cluster NGC~1904.
Figure 10 illustrates this phenomenon with data for the globular clusters
M~5, M~15, M~92, NGC~6397 and NGC~288 in addition to the data of this paper.
As can be seen there} the mass distribution of the BHB stars differs
considerably from those of the sdB stars analysed here. The fact that the
mean masses of BHB stars in globular clusters
are significantly and systematically
too low has already been noted in papers I and II. Most of the possible
explanations of this finding (see papers I and II), however, are
ruled out by the mass distribution
of the sdB stars presented here: Non-solar helium abundances, peculiar
metallicities due to diffusion, helium stratification, systematic errors
in \logg\ and/or distance moduli should affect the sdB stars in the
same way as the BHB stars.
\item[(iv)] Further insight into the nature of BHB and EHB stars may come
from their radial distribution.
We therefore investigated the radial distribution of all stars in Buonanno et
al. (\cite{buca86}) that are fainter than V~=~\magpt{13}{4} and bluer than
B$-$V~=~\magpt{-0}{2}. This sample, consisting of 214 stars, was divided
into two groups: BHB stars with V~$<$~16 (142), and EHB
stars with V~$\ge$~16 (70), which combines all sdB stars.
We then counted the stars in radial bins of 50\arcsec\
 and normalized the numbers to the total numbers of stars in each group
as well as to the area covered by each bin.
The resulting radial distributions show no significant difference,
supporting a similar evolutionary history
and mass distribution, in contrast to the situation described above.
\end{itemize}
While the idea that there is something wrong with the model atmospheres,
leading
to too low values of \logg\ for the temperature range between 10000 K and
20000 K sounds rather intriguing and would also explain the loci of the
BHB stars in the \logt -\logg -diagram (cf. Fig. 9 of Paper~I), it could not
explain the findings of Paper~II, where the physical parameters of the stars
fit very well to theoretical expectations, but result nevertheless in too low
masses. The remaining possibility, i.e. that the BHB masses {\em are}
indeed that low
puts severe problems to stellar evolutionary models, since up to now no
scenario is known to produce stars in that temperature range with masses
{ of about 0.3~\Msolar .}

\begin{figure}
\includegraphics{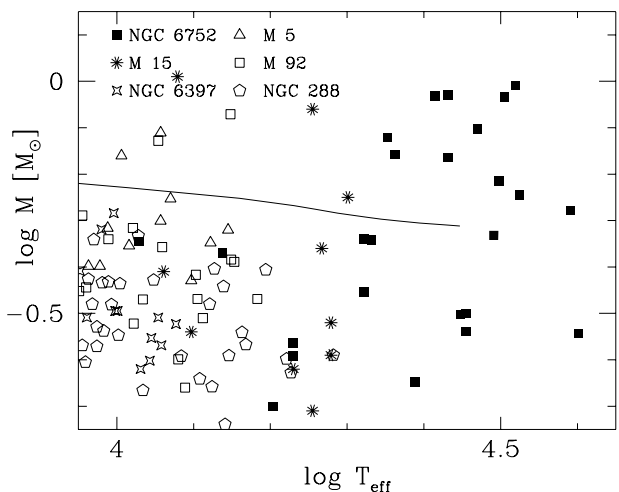}
\vspace{6.3cm}
\caption[]{The resulting masses plotted in Fig. 9 compared to masses of BHB
stars in other cluster. The BHB data are taken from Paper I (M 15), II
(NGC~6397), and
Crocker et al. (1988, M~5, M~92, and NGC~288)}
\end{figure}

\acknowledgements

We want to thank the staff of the ESO La Silla observatory
for their support during our observations and especially Drs. D. Baade and
A. Zijlstra for their assistance with our calibration problems.
We are grateful to Dr. R.C. Bohlin for his help with
calibration of the IUE data, to Drs. U. Hopp and M. Rosa
for their assistance with the wavelength calibration,
and to Dr. R. Napiwotzki for his help with B~852.
Thanks go also to an anonymous referee for valuable advice,
to Dr. C.E. Corsi for the NGC 6752 data, to Dr. C. Cacciari for
important discussions, and to Dr. R. Saffer for the permission to use
his fit routines.
SM acknowledges support by the DFG (grant
Mo 602/6-1), by the Alexander von Humboldt-Foundation, and
by the director of STScI, Dr. R. Williams, through a DDRF
grant.
This research has made use of the SIMBAD data base, operated at CDS,
Strasbourg,
France.

\end{document}